%% file: bare_jrnl.tex
\documentclass[journal,onecolumn]{IEEEtran}
\usepackage{amsmath,amsfonts}
\usepackage{algorithmic}
\usepackage{algorithm}
\usepackage{array}
\usepackage{textcomp}
\usepackage{stfloats}
\usepackage{url}
\usepackage{verbatim}
\usepackage{graphicx}
\usepackage{cite}
\usepackage{enumitem}
\usepackage{xcolor}
\usepackage{balance}

\usepackage{LocalConfig}
\usepackage{linguex}
\usepackage{comment}
\usepackage{multirow}
\usepackage{times}
\usepackage{booktabs,chemformula}
\usepackage[T1]{fontenc}
\usepackage{lipsum}

\usepackage[usestackEOL]{stackengine}
\edef\tmp{\the\baselineskip}
\setstackgap{L}{\tmp}

\usepackage{pifont}
\usepackage{amssymb}
\newcommand{\xmark}{\ding{55}}%
\ifCLASSINFOpdf
\else
\fi
\begin{document}
%
\title{Web Application Weakness Ontology Based on Vulnerability Data}
%
%
%

 \author{
 \IEEEauthorblockN{\small Onyeka Ezenwoye$^*$$^1$, Yi Liu$^!$}\\
\IEEEauthorblockA{\small $^*$ Augusta University, Augusta, GA 30912, USA, { \tt oezenwoye@augusta.edu } \\
          $^!$ University of Massachusetts Dartmouth, Dartmouth, MA 02747, USA, { \tt yliu11@umassd.edu}}

}

\maketitle
\input{s0_Abstract}

%
\IEEEpeerreviewmaketitle

\input{s1_Introduction}

\input{s4_Weaknesses}

\input{s2_Methodology}
\input{s5_RelatedWork}

\input{s6_Conclusion}

\input{s7_Appendix}


\ifCLASSOPTIONcaptionsoff
  \newpage
\fi



%
\bibliographystyle{IEEEtran}
\bibliography{onyeka}

\end{document}

%% file: s0_Abstract.tex
\begin{abstract}
Web applications are becoming more ubiquitous. All manner of physical devices are now connected and often have a variety of web applications and web-interfaces. This proliferation of web applications has been accompanied by an increase in reported software vulnerabilities. The objective of this analysis of vulnerability data is to understand the current landscape of reported web application flaws.
Along those lines, this work reviews ten years (2011 - 2020) of vulnerability data in the National Vulnerability Database. Based on this data, most common web application weaknesses are identified and their profiles presented. A weakness ontology is developed to capture the attributes of these weaknesses. These include their attack method and attack vectors. Also described is the impact of the weaknesses to software quality attributes. Additionally, the technologies that are susceptible to each weakness are presented, they include programming languages, frameworks, communication protocols, and data formats.
\end{abstract}


%% file: s1_Introduction.tex
\footnotetext[1]{corresponding author}

\section{Introduction}
\label{s:Introduction}
All manner of applications and devices are now connected with web-enabled interfaces for remote interaction. Software tools exist to allow individuals at various levels of technical expertise to quickly couple together disparate software components and technologies to spin-up a web application. With this proliferation however comes the associated software vulnerabilities. These vulnerabilities occur in many different forms and at various layers of the application stack. To help address the prevalence of these vulnerabilities, it is important to understand their nature.
The National Vulnerability Database (NVD) maintains a collection of vulnerabilities that exist in everyday software~\cite{NVD2021}.
The NVD was initiated in 2002 by the National Institute of Standards and has since cataloged tens of thousands of vulnerabilities over the years. These vulnerabilities are instances of discovered security-related faults in various software.

For this work, we leverage the information in the NVD to gain some insight into the vulnerabilities that exist in Web applications. We create a taxonomy of Web application weaknesses by taking a contemporary view of their associated vulnerabilities.
Addressing the issues of software vulnerabilities require some understanding of their characteristics. This insight is useful for stakeholders from industry and education alike in the push to root out security faults. In 2006, the United States Department of Homeland Security warned that software should be developed with security in mind during every phase of the software development lifecycle~\cite{Goertzel2006}. To accomplish this, software practitioners require the necessary tools at various stages of the process.

Early in the development process, threat modeling is needed to understand the vulnerability landscape and refine the system's requirements accordingly~\cite{Tuma2018, Writing2002}. Throughout the process, checks are needed to ensure that the software is developed in a manner that is consistent with those requirements. 
Inspection is such a verification technique that can expose problems at various stages of the software process. Checklists, used during inspections, are a powerful tool for fault detection especially when used early in the software process~\cite{OOSE2009}. The usefulness of checklists is predicated on knowing common problem areas. Checklists are meant to evolve over time as new problems crop up and this is true for various professions from home inspectors to airline pilots. Software applications are often constructed by coupling together existing components with a multitude of technologies from various vendors. The composite software will inherit any flaws that exist in the components. So, understanding the nature of the existing vulnerabilities is important. This is particularly true for web applications that could have many remote components and actors~\cite{Cox2019}. We define web applications as those that are inherently client-server architecture and who's subsystems interact over the internet utilizing a request-response pattern~\cite{Ivaki2015}. Web applications leverage existing web infrastructure, standardized communication protocols and are platform-independent~\cite{guinard2011, hadley2006, guinard2011web}.

With this work we answer the following research questions:
\begin{enumerate}[label=(\Alph*)]
    \item How many reported vulnerabilities in the NVD, over the last ten years (2011 to 2020), impact web application software?
    \item What are the most common weaknesses in the reported vulnerabilities for web applications?
    \item What attack methods are utilized to exploit each weakness?
    \item Which attack vectors were leveraged in exploiting each weakness?
    \item Which technologies are susceptible to each weakness?
\end{enumerate}
Quite a number of works have looked at the problem of understanding and classifying computer system vulnerabilities. We discuss these works in Section~\ref{s:RelatedWork}. None of these existing works take a contemporary view at web application vulnerability data, nor do they address any of the research questions above. Besides, some of these works are merely proposals that are not based on empirical analysis. With the work presented here, we address the above questions. We present a weakness classification mechanism with various components, including typical attack methods, attack vectors, and impacted technologies. We identify the most susceptible programming languages, frameworks, communication protocols, in addition to data and media formats. We also briefly describe each of the weaknesses and some ways to mitigate them. With this work, we intend to give an updated view on the attributes of web application weaknesses. This is to better inform stakeholders in the push to improve software processes and training for engineers.

The rest of this paper is structured as follows; Section~\ref{s:Weaknesses} provides some background on vulnerabilities and the 11 types of weaknesses discussed in this paper. We describe our methodology and the taxonomy hierarchy in Section~\ref{s:methodology}. A discussion on related work can be found in Section~\ref{s:RelatedWork} with concluding remarks in Section~\ref{s:Conclusion}.

%% file: s4_Weaknesses.tex
\section{Background}
\label{s:Weaknesses}

This work is an extension of previous work on vulnerability data analysis by the authors~\cite{SoftwareType-SEKE-2020, Ezenwoye2022, Lee21}. These previous works leverage software vulnerability data from the National Vulnerability Database (NVD)~\cite{NVD2021}. They looked at the types of vulnerabilities that impact different types of software, vulnerability risk factors, and ways to mitigate them. This section provides some background on the types of vulnerabilities that are discussed in this paper. This material is useful as these types of vulnerabilities are frequently mentioned. Readers who are familiar with this background may skip this section.

The NVD  captures security vulnerabilities in a list known as Common Vulnerabilities and Exposures (CVE)~\cite{CVEProgram}. Each CVE entry contains several attributes including a unique \emph{identifier}, the \emph{name} of the affected software product, the software \emph{vendor}, a \emph{description} of the vulnerability, and the \emph{date} of the report.
Each entry is also associated with a type of weakness in a Common Weakness Enumeration (CWE). CWE is a list of software and hardware weakness types (e.g., Buffer Overflow)~\cite{CWE}. So, each vulnerability (CVE), which is an instance of a reported software fault, is associated with a specific weakness (CWE). Each CWE item has a unique name and identifier. For instance, the identifier for Buffer Overflow is CWE-119. From here on, we refer to CVE entries as \emph{Vulnerabilities} and CWE items as \emph{Weaknesses}. Below are further detail about the 11 weaknesses discussed in this paper. The goal is to briefly introduce each weakness and provide some context for the rest of the material.

\subsection{Cross-site Scripting (XSS)}

This weakness results from the consumption of malicious input, which is then incorporated into the application's output. An actor can exploit a web browser's ability to run embedded code by inserting malicious web scripts (E.g., JavaScript, HTML) into system data. For instance, malicious script could be inserted into the payload of a URL, form data, email, or file. The malicious script is then consumed and inadvertently executed by the browser. The malicious code, which is disguised to execute within the security context of a trusted domain, redirects the actor or actor's data to an untrusted domain~\cite{Exploiting2004, Howto2006}. Several techniques for preventing XSS exist and include avoiding the use of user-supplied input (E.g., request parameters) in application logic, validating application output, applying the HTTPOnly flag to cookies to make them invisible to client-side code, and specifying an encoding type in HTTP response headers~\cite{stuttard2008, sullivan2011}.


\subsection{SQL Injection}
 SQL (Structured Query Language) statements are sometimes constructed dynamically within a program using input data from the actor. For instance, input data could be used to construct a query to authenticate the actor or conduct a search for products. SQL injection is a type of injection attack that occurs when an actor can insert portions of SQL statements into a program's input data. The malicious input is designed to exploit weak data validation and poorly constructed dynamic SQL queries to gain access to the database.
 Similar to XSS, mitigating SQL injection includes data validation, sanitization, and avoiding the use of user-supplied input in SQL queries. Other techniques include making use of stored procedures rather than dynamic queries, and using queries than are resilient to SQL injection~\cite{sullivan2011}.


\subsection{Exposure of Sensitive Information}
This weakness manifests in various forms.
The vulnerabilities associated with this weakness include overly descriptive error messages (E.g., stack traces), insecure cookies, information logs (including caches), and other system artifacts (E.g., client-side code) that reveal information such as IP addresses, user credentials, document links, and database connection strings. An actor could glean useful information about the system and utilize the information to refine a future attack. Error messages from login attempts and password resets can also be exploited in this manner.
%
Revealing sensitive information in URL parameters also falls within this weakness. The vulnerabilities reported here included, not properly implementing access controls which allows actors to obtain sensitive information through direct requests. Others are, not encrypting network traffic, and in some cases, submitted passwords have been included within HTTP response messages. Also reported were username and password fields that revealed user credentials through auto-complete. Mitigating this weakness requires a very broad approach including using generic error messages, effective access control, strong data encryption, and minimizing client-side information disclosure~\cite{stuttard2008, TheArt2006, Secure2007}.

\subsection{Cross-Site Request Forgery (CSRF)}

This weakness exists when an application is unable to correctly determine if a client request is intentionally sent. A lack of proper verification of requests could be exploited by an attacker to trick a user to submit a malicious request, for instance by clicking a link (URL) while logged into a system. This request thus leverages the context of the user's valid session. The forged requests could be used to carry out malicious actions on behalf of the attacker. Malicious actions could include creating a user account or modifying a user's credentials. 
This weakness occurs mainly in applications that utilize HTTP cookies as the only means for tracking sessions. Mitigating CSRF centers around having proper session token checks which might include using unique request tokens and inserting tokens into hidden form fields~\cite{stuttard2008, sullivan2011, Secure2007}.

\subsection{Improper Access Control}

This weakness relates to failure to correctly \emph{authenticate} the identity of an actor and failure in correctly determining whether the actor is \emph{authorized} to access some functionality or information. This could result from the improper management of session identifiers as well and lack of consistent enforcement of authentication and authorization across different layers of the architecture. With that, an actor that is able to bypass authentication measures at the presentation layer may find access control that incorrectly authorizes access to data and functionality~\cite{Enterprise2014, TheArt2006}.
Some aspects of this weakness relate to improper management of user credentials.
A few of the vulnerabilities reported here include poor password management that allows an actor to change a password without knowing the existing password. Poor password policies have also allowed for successful brute-force password guessing. Mitigating this weakness must include the implementation of appropriate access control patterns (E.g., role-based access control) as well as strong policies for managing user credentials.

\subsection{Improper Input Validation}

This weakness involves the inadequate validation of data that originates from outside the trust boundary of an application. By inadequately validating, the application places undue trust on consumed data by discounting the possibility of malicious data. Consuming malicious data is the basis for the injection attack method and can cause an application to enter an erroneous state that leads to various other weaknesses. The vulnerabilities reported under this weakness include, not properly handling malicious server requests which could contain manipulated session data, certificates, or other invalid request parameters. Others are, not validating the size and content of uploaded files, and not validating form. This weakness could be mitigated by identifying all data flows that originate from outside the boundary of a system component (including packet data). All incoming data need to be validated~\cite{Writing2002, Enterprise2014, Secure2007}.

\subsection{Path Traversal}

An application may utilize user-controlled data to generate a file path to the system directory.
An actor could exploit this behavior to gain access to an unauthorized location in the directory tree.
Through path traversal, an actor could access system resources that permits them to impact the reliability and availability of the application as well as violate the integrity and confidentiality of system data~\cite{TheArt2006, Secure2007}. The vulnerabilities reported under this weakness include allowing an actor to access or delete files via a malicious URL. For instance, a malicious URL could be crafted by replacing portions of a legitimate URL with a dot-dot-slash sequence (E.g., ../../)
Mitigating path traversal could include not allowing user-controlled data as input to the file system, otherwise such data needs to be thoroughly sanitized of suspicious characters~\cite{stuttard2008}

\subsection{Buffer Overflow}

In this weakness, memory location that is outside the bounds of a buffer can be accessed if proper bounds checks are absent. An actor can exploit this vulnerability by supplying data that causes the overflow to occur. Access could be gained to other parts of the program's memory which can cause the application to behave as unintended~\cite{Secure2007}. The vulnerabilities reported under this class include the insertion of invalid data through office productivity documents, XML files, email attachments, and crafted websites with malicious code. Preventing this weakness requires appropriate validation of all forms of input data, including the length of URL strings~\cite{stuttard2008}.

\subsection{Unrestricted Upload of File}
This weakness relates to vulnerabilities that allow an actor to specify a malicious file that is consumed by the application. The file is then either automatically processed by the application or directly accessed by the actor. Mitigating this weakness requires that both a file's payload and type (extension) are adequately validated. Suitable whitelists or blacklists could be used to filter allowable types. Also, access control measures should be used to prevent unauthorized uploads.

\subsection{Code Injection}
In this weakness, the actor is allowed to insert code snippets into a system and thereby alter its behavior. The vulnerabilities reported for this weakness include the uploading of executable files and insertion of source code via serialized objects, email attachments, media files, and ZIP archives. Steps to mitigate this weakness should include adequate validation and sanitization of data. Additionally, access control measures should be in place to prevent access to application features and components that might be susceptible to code injection.

\subsection{Improper Authentication}
This weakness involves the incorrect implementation of security-related features (E.g., configuration setting and session management) that allows an actor to bypass authentication. The NVD notes that Improper Authentication is related to (child-of) \emph{Improper Access Control}~\cite{CWE-287}.
Vulnerabilities reported here included enabling auto-complete for authentication fields, not requiring authentication for requests, and improper sanitization of authentication input.

%% file: s2_Methodology.tex
\section{Methodology and Results}
\label{s:methodology}

To understand the weaknesses that are associated with web applications, empirical evidence is needed. Along this vein, we decided to review reported software vulnerabilities in the NVD.
The NVD contains several thousands of vulnerabilities across several thousands of software products. With NVD data, some generalized vulnerability and weakness trend analysis is possible~\cite{VulnViz2021, CVEDetails2021, Homaei2017}. Although vulnerability entries in the NVD contains attributes such as vendor and product, the data does not identify the software type (e.g., Operating System). For instance, a vulnerability for Windows 10 would identify Window 10 without identifying its type as operating system. So, to identify web application vulnerabilities, we needed to classify each vulnerability by the software type of the named product. Identifying software type is important for such empirical study as important characteristics of software do vary by their type. Determining the application type is important in performing analysis that is specific to the application domain~\cite{Forward2008}.

\begin{figure}[!htb]
\centering
\includegraphics[width=3.49in]{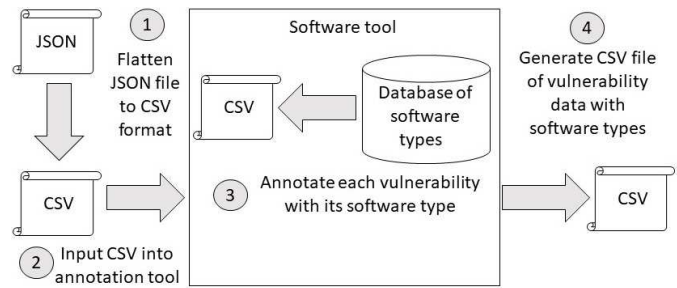}
\caption{Vulnerability data annotation process}
\label{f:conversionProc}
\end{figure}

In previous work, we classified NVD vulnerability data into seven software types, namely; \emph{web application, utility, server, operating system, browser (web), framework}, and \emph{middleware}~\cite{SoftwareType-SEKE-2020, Ezenwoye2022}.
The process of classifying the software type is illustrated in Figure~\ref{f:conversionProc}. The NVD provides a feed of yearly vulnerability data~\cite{NVDData2021}. The data which is available in JSON (JavaScript Object Notation) format does not identify the software type for each vulnerability, so we first annotate each vulnerability with a software type. Our four-step annotation process is described below:

\begin{enumerate}
    \item The JSON data format is first converted to a CSV (Comma-Separated Values) format. Repeated entries of the same vulnerability (with the same identifier) are removed.
    \item The CSV list of vulnerabilities is fed into a software tool that we specifically developed for annotating vulnerabilities with a software type.
    \item A software type is inserted as an attribute of each vulnerability. The type is identified from the combination of the vendor and product name (e.g., Microsoft and Windows 10). We created a database of software types, which is utilized to make this match.
    \item A CSV file of vulnerabilities is generated. Each vulnerability is now associated with one of the six software types previously mentioned. 
\end{enumerate}

Our analysis is based on reviewing the CSV vulnerability data. This manual review includes reading the description provided for each reported vulnerability. The vulnerability data in CSV format is viewed using Microsoft Excel spreadsheet. Pivot tables~\cite{dierenfeld2012} are used to filter the data for analysis. This method of manual review is tedious. A scalable approach to review of vulnerability descriptions would require automated tools that employs natural language processing and machine learning techniques.

\subsection{How many reported vulnerabilities in the NVD, over the last ten years (2011 to 2020), impact web application software?}
\begin{figure}[!htb]
\centering
\includegraphics[width=3.15in]{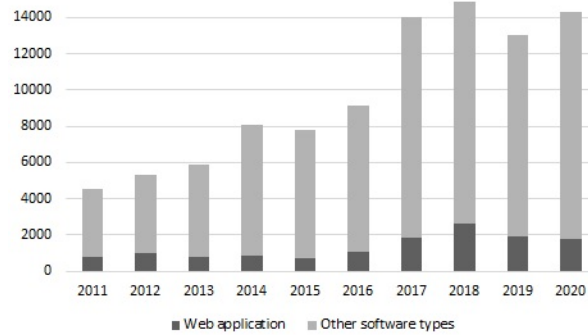}
\caption{Web application vulnerabilities vs all vulnerabilities, 2011-2020}
\label{f:vulnTotals}
\end{figure}

After annotating the vulnerability data with software type, we focused on a decade's worth of vulnerabilities (2011 to 2020).
We then picked those of the web application type for which there were a total of 13,346 vulnerabilities. Figure~\ref{f:vulnTotals} shows the distribution of web application vulnerabilities in the review period. Although there are 7 types of applications in our classification. We focus on web applications because software is increasingly connected. This has led to the blurring of lines between application types. Contemporary applications increasingly have elements that are web-based~\cite{michahelles2015, guinard2011}. Our data shows that the most vulnerabilities were reported in 2018 (14,863). Of these, 18\% (2,644) were associated with web applications. Over each of the ten years, the portion of web application vulnerabilities ranged from 9\% (704) in 2015 to 19\% (979) in 2012. Our analysis showed that over the ten years, operating system vulnerabilities accounted for the most vulnerabilities (23,567), while utility software vulnerability was second (20,790). Server software was third (16,561), Frameworks were fourth (16,099), web application was fifth. The rest were, Browser and Middleware (5,538 and 983, respectively)

\subsection{What are the most common weaknesses in the reported vulnerabilities for web applications?}

After determining which vulnerabilities are attributed to web application software, we then determine the most common weaknesses in that group of vulnerabilities.
Thus, we reviewed the web application vulnerabilities to identify the most prevalent weaknesses. Our criteria for prevalence is based on a \emph{weight} metric. The weight is the frequency of occurrence of a weakness in a group of vulnerabilities. The threshold for most common is any weakness that has a weight of 1 or more (frequency of 1\% or greater). By this metric, 11 weaknesses met the threshold. Table~\ref{t:MostWeaknesses} shows the 11 most common weaknesses and their respective weights. Our analysis shows a total of 127 identified weakness over the 13,346 vulnerabilities for web applications.
Of the 127 weaknesses, these 11 weaknesses accounted for just over 70\% of all vulnerabilities for web applications. As can be seen from Table~\ref{t:MostWeaknesses}, Cross-Site Scripting is the most common weakness for web applications. In Section~\ref{s:Weaknesses}, we briefly describe each weakness and present further details on our findings.

\begin{table}[!htb]
\renewcommand{\arraystretch}{1.1}
\caption{Most Common Weaknesses for Web Application (2011 - 2020)}
\label{t:MostWeaknesses}
\centering
\begin{tabular}{|l|l|c|}
\hline
\textbf{ID} & \textbf{Name} & \textbf{Weight}\\
\hline
CWE-79 & Cross-Site Scripting (XSS)& 27.6\\
\hline
CWE-89 & SQL Injection & 9.9\\
\hline
CWE-200 & Exposure of Sensitive Information & 7.0\\
\hline
CWE-352 & Cross-Site Request Forgery (CSRF) & 5.8\\
\hline
CWE-284 & Improper Access Control & 5.2\\
\hline
CWE-20 & Improper Input Validation & 4.5\\
\hline
CWE-22 & Path Traversal & 3.3\\
\hline
CWE-119 & Buffer Overflow & 2.3\\
\hline
CWE-434 & Unrestricted Upload of File & 2.0\\
\hline
CWE-94 & Code Injection & 1.8\\
\hline
CWE-287 & Improper Authentication & 1.0\\
\hline
\end{tabular}
\end{table}

\begin{figure*}[!htb]
\centering
\includegraphics[width=5.0in]{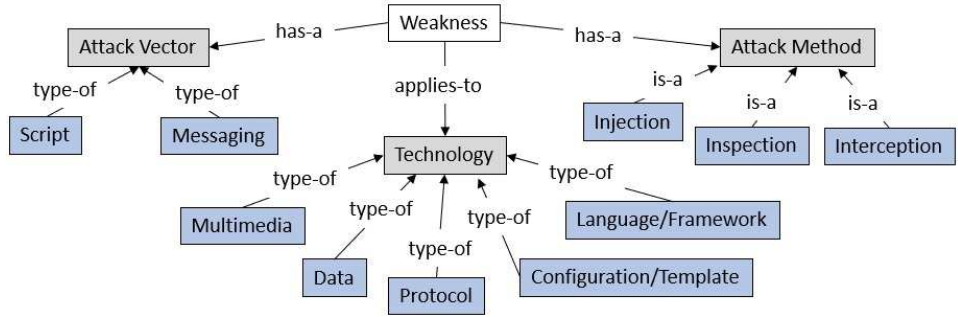}
\caption{Weakness attribute ontology}
\label{f:weakness_ontology}
\end{figure*}

The rest of the research questions refer to specific attributes of each of weaknesses. This requires that the vulnerabilities associated with each weakness in Table~\ref{t:MostWeaknesses} be analyzed to identify these attributes. Each vulnerability was reviewed (recall that each vulnerability is associated with a weakness).
Our analysis of the vulnerabilities informed the classification of the attributes of each weakness.
Figure~\ref{f:weakness_ontology} illustrates the relationship between a weakness and 3 attributes (\emph{Attack Vector, Attack Method,} and \emph{Technology}). Each attribute has a set of children. We use the \emph{type-of} relationship where the child is not atomic, i.e., it has other subtypes. We use the \emph{is-a} relationship where the child is atomic. In our analysis, each vulnerability was reviewed to determine its specific attributes. The ontology is based on the answers to the research questions posed here. It summarizes the results the analysis of contemporary vulnerability data to understand the various aspects of web application weaknesses. This understanding existing vulnerabilities is important to inform related endeavors in threat modeling and security requirements engineering. In these endeavors, there are increasing needs for automated tools to assist in tackling the growing vulnerability landscape~\cite{Tuma2018,shostack2014,MELLADO2010,Souag2016}. We provide further details of these weakness attributes as we address each research question in the sections below. In Section~\ref{s:evaluation}, we offer a comparison with similar works that include some sort of vulnerability classification.

\subsection{What attack methods are utilized to exploit each weakness?}
 
We define \emph{attack method} as the technique that is utilized by an attacker (malicious actor) to exploit a weakness. To determine the attack method for each weakness, all vulnerability (cve) data were reviewed. Although the level of detail for the descriptions did vary, there was at times enough information to determine attack methods. Our review identified three types of attack methods, namely, \emph{Injection, Inspection, and Interception}. Each attack method is described as follows:

\begin{enumerate}
    \item \emph{Injection}: In this method, the attacker inserts some form of malicious data into the control flow of an application. An example of an injection attack is uploading an image file that has been compromised with malicious code. Other forms of this attack include the injection of commands into system, to perform tasks on behalf of the attacker. This could be accomplished through maliciously crafted scripts or action by the attacker to access a functionality (or issue requests) for which they are not authorized. Mitigating injection attack requires that undue trust is not placed on data that originates from outside the trust boundary of an application. Such data needs to be validated and invalid data be discarded. Implementation of adequate access control measures are also useful for mitigating injection attacks.
    \item \emph{Inspection}: In this method, the attacker is able to observe sensitive information about the application for which they are not authorized. For instance, an attacker could review source code and other system artifacts (e.g., client-side views, logs and configuration files) to find login credentials, IP addresses, or other information that reveal sensitive implementation detail. Often, an inspection-based vulnerability is the precursor to a future attack as the attacker is able to leverage the observed information to exploit other weaknesses. To mitigate this weakness, care must be taken to scrub exposed system artifacts of sensitive information. Exposed system artifact are those that a put outside of a system's trust boundary where they are easily accessible to an attacker. Encryption should be considered for certain artifacts and caches cleared when no longer needed. Also, care must be taken to see that error messages are generic and do not expose sensitive information to an attacker. Examples of inspection-related vulnerabilities are ~\cite{CVE-2011-0774} and~\cite{CVE-2012-0792}. 
    \item \emph{Interception}: In this method, the attacker is able to view or modify sensitive data in the application's network traffic. This is typically known as a \emph{man-in-the-middle} attack as the attacker is able to intercept the traffic between a sender and receiver. The attacker takes advantage of absent or poor encryption of network messages. CVE-2015-0943~\cite{CVE-2015-0943} is an example of a reported vulnerability that utilizes this attack method.
\end{enumerate}

Figure~\ref{f:weakness_ontology} illustrates the relationship between attack method and weakness.
Each vulnerability is associated with a weakness and one or more attack methods. For instance, a reported vulnerability could reveal that login credentials were discovered in an application's source code. For this vulnerability, the associated attack method would be Inspection, while the weakness is \emph{Exposure of Sensitive Information}. This relationship allows us to identify all attack methods for each weakness.
The association between all the weaknesses and attack methods is captured in Table~\ref{t:MethodVectorMapping}. It shows the most common web application weaknesses and their observed attack methods. Based on our analysis, \emph{Injection} is the attack method that is utilized across all types of weaknesses. Also, \emph{Exposure of Sensitive Information} is the only weakness that was found to be susceptible to all three attack methods. This mapping is based on the vulnerability data reviewed. It is conceivable the some weaknesses are susceptible to any of the three attack methods.

\begin{table*}[!htb]
\NewDocumentCommand{\rot}{O{45} O{1em} m}{\makebox[#2][l]{\rotatebox{#1}{#3}}}%
  \centering
  \caption{Weaknesses and associated attack methods and attack vectors (script \& messaging)}
  \label{t:MethodVectorMapping}
  \begin{tabular}{@{}|l|c|c|c|c|c|c|c|c|c|c|c|c|c|@{}}
    \hline
    \textbf{Weakness} & \multicolumn{3}{ c |}{\textbf{Method}} & \multicolumn{4}{ c |}{\textbf{Script}} & \multicolumn{6}{ c |}{\textbf{Messaging}}\\ 
\cline{2-14}
    & \rot[90]{\textbf{Injection}}
    & \rot[90]{\textbf{Interception}}
    & \rot[90]{\textbf{Inspection}}
    & \rot[90]{\textbf{Form Data}}
    & \rot[90]{\textbf{Source Code}}
    & \rot[90]{\textbf{File Data}}
    & \rot[90]{\textbf{Cookie Data}}
    & \rot[90]{\textbf{E-mail Message}}
    & \rot[90]{\textbf{RPC Request}}
    & \rot[90]{\textbf{HTTP Request}}
    & \rot[90]{\textbf{STUN Request}}
    & \rot[90]{\textbf{DNS Request}}
    & \rot[90]{\textbf{AJAX Request}}\\
    \hline
    Cross-Site Scripting (XSS) &x&&& x & x & x & x & x & x & x &  &  & x\\
    \hline
    SQL Injection &x&&& x & x & x & x &  & x & x &  &  & x\\
    \hline
    Exposure of Sensitive Information &x&x&x& x & x & x & x &  & x & x  & x &  & x\\
    \hline
    Cross-Site Request Forgery &x&&& x &  & x & x &  & x & x &  &  & x\\
    \hline
    Improper Access Control &x&&& x & x & x & x & x & x & x &  &  & x\\
    \hline
    Improper Input Validation &x&&& x & x & x & x & x & x & x &  & x & x\\
    \hline
    Path Traversal &x&&& x & x & x &  &  &  & x &  &  & x\\
    \hline
    Buffer Overflow &x&&& x & x & x &  & x & x & x &  &  &\\
    \hline
    Unrestricted Upload of File &x&&& x & x & x &  &  & x & x &  &  & x\\
    \hline
    Code Injection &x&x&& x & x & x & x &  & x & x &  &  & x\\
    \hline
    Improper Authentication &x&&& x & x & x & x &  & x & x &  & x & x\\
    \hline
  \end{tabular}
\end{table*}

\subsection{Which attack vectors were leveraged in exploiting each weakness?}

\emph{Attack vector} can be broadly defined as a pathway through which an attacker exploits a system~\cite{Simmons14}. For this work, we view those pathways as the \emph{part of} the application that serves as the delivery mechanism or is the target of an attack that seeks to exploit a weakness. This attribute has two subtypes, namely, \emph{Script} and \emph{Messaging}. As illustrated by Figure~\ref{f:weakness_ontology}, these attributes are not atomic and therefore have their own subtypes. Table~\ref{t:MethodVectorMapping}\footnote{See appendix for explanation of acronyms} captures the association between the most common web application weaknesses and their attack vectors. This is based on our analysis of vulnerabilities that occurred over the review period of 2011 to 2020. Script and Messaging have four and six subtypes, respectively. A keyword-based search of the all vulnerability data was used to determine the association between weaknesses and vector. Example keywords were "form", "upload", "cookie", "ajax", and "rpc". The keywords were chosen based on the review vulnerability descriptions After a search determines the presence of a keyword in a vulnerability, the descriptions is reviewed to determine whether the keyword is used in the correct context. That is, to determine if the keyword is an attack vector for the associated weakness. CVE-2020-12432~\cite{CVE-2020-12432} is an example of a vulnerability that leverages file data to exploit XSS. The vulnerability description includes the keyword "upload". We further the identified attack vectors below:

\subsubsection{Script}
Script represents application artifacts that occur in some form of textual or character encoding.  Script has four subtypes which are as follows:
\begin{enumerate}[label=\roman*.]

    \item \textit{Form Data}: 
    Malicious data could be inserted into a form field to exploit weaknesses.
    This attack vector is typically used to carry out injection-based attacks where the attacker injects data that exploits a weakness. With this attack vector, the attacker takes advantage of the lack of input data length limits, or poor validation and sanitization of input data. CVE-2019-9714~\cite{CVE-2019-9714} is an example of a vulnerability that is associated with form data and exploits XSS weakness. As Table~\ref{t:MethodVectorMapping} show, we found form data as an attack vector for all types of weaknesses.
    \item \textit{File Data}: Various file formats can be manipulated by an attacker and used to exploit a weakness.
    This is typically accomplished through a file upload (or attachment) function. An attacker can leverage these functions if the application lacks adequate access control, proper checks for file size limit, and validation for file format (both file content and file extension)~\cite{CVE-2020-13241}. Examples of utilized file formats include images and productivity (E.g., Microsoft Word).
    XML data can be susceptible to various types of attacks such as XML eXternal Entity (XXE) injection. \emph{XXE injection} leverages the ability of XML documents to reference data that exists outside the document~\cite{CVE-2017-12216}. For instance, a manipulated URL (Uniform Resource Locator) in a JSON (JavaScript Object Notation) file could be used to trick the browser to download and execute a malicious file, in an attack known as \emph{reflected file download}~\cite{CVE-2017-18123}. In some cases, remote and local inclusion of configuration, template, data, and source files have been leveraged to compromise a system~\cite{CVE-2017-1000496}. Table~\ref{t:MethodVectorMapping} shows that File Data is a common attack vector for all types of weaknesses.
    \item \textit{Source Code}: An attacker could leverage languages such as SQL, PHP (Hypertext Preprocessor), and JavaScript to craft malicious code snippets that are then injected into the application via some other attack vector such as Form Data or File Data~\cite{CVE-2020-11807, CVE-2017-12979}. At times, the very nature of some file formats allows for them to be injected with malicious source code. For instance, PHP code can be embedded in some image files~\cite{CVE-2017-17727}. An attacker could also glean useful information about an application by inspecting different forms of its implementation artifacts such as client code, configuration files and local data cache~\cite{CVE-2016-3002, CVE-2015-1951}.
    \item \textit{Cookie Data}: A cookie is a name-value pair data structure that is part of the HTTP protocol. It is used to exchange application-context information between the server and the client. The types of information exchanges might include session identifier, expiration date, and a secure flag that determines if the cookie is submitted only in HTTPS. In some cases, an application may extract information in the HTTP request parameters, such as user preferences, and place that information within a cookie in the response message. An attacker may leverage this behavior to carry out a cookie injection attack by crafting a malicious request that would insert a vulnerability into cookie data~\cite{CVE-2015-5687}.
    The susceptibility of cookies highlights the importance of encryption, validating cookie data, sanitizing user supplied data, and utilizing good server-side session management.
   \end{enumerate}
   
\subsubsection{Messaging}
Messaging represents communication mechanisms for data interchange over a network.
These mechanisms can be used as attack vectors to exploit weaknesses. We identify six subtypes of this attribute and describe them below:
\begin{enumerate}[label=\roman*.]
    \item \textit{E-mail Message}: This attack vector relates to vulnerabilities that involve the exchange of e-mail messages. Weaknesses have been exploited by embedding malicious code into e-mail messages~\cite{CVE-2014-1695}. Also possible are account activation vulnerabilities (related to e-mail verification) that allow for accounts to be created by the use of tokens in e-mail messages~\cite{CVE-2017-8385}. Other vulnerabilities include e-mail header injection in which the addressing component of an e-mail message can be manipulated with code to result in outbound spam messages. We also observed some vulnerabilities where an attacker injects executable code into the attachment in an e-mail message~\cite{CVE-2018-15408}. Table~\ref{t:MethodVectorMapping} shows the associated weaknesses to this attack vector.
 
    \item \textit{HTTP Request}: The Hypertext Transfer Protocol (HTTP) is a request-response protocol for client-server interaction~\cite{HTTP}. HTTP requests are utilized to exploit multiple types of weaknesses. For instance, an authenticated actor could compose a request to a path that gives them access to unauthorized functionality or data, thereby escalating their privilege and exploiting an \emph{Improper Access Control} weakness~\cite{CVE-2019-11489}. CVE-2013-2713~\cite{CVE-2013-2713} is another example of various forms of weaknesses that have been exploited using maliciously crafted requests.
    %
    A handful of vulnerabilities involved the exposure of sensitive information in HTTP responses. The exposed information has included submitted passwords and private IP addresses~\cite{CVE-2011-4737}. Extraneous information needs not be included in responses, especially when unobscured.

    HTTP messages contain metadata (header) that includes information such as the host address, browser name, user language, and character encoding. The message header and payload are separated by a combination of carriage return (\%0d) and line feed (\%0a) characters, known as \emph{CRLF}~\cite{stuttard2008}. HTTP headers are susceptible to manipulation where an attacker leverages improper sanitization of user supplied data to inject vulnerabilities into the message. Attackers have carried out CRLF injections through a crafted request that would create a malicious header or payload and exploit various types of weaknesses~\cite{CVE-2011-4203}. An example is the \emph{"Referer" HTTP header} attack which is used to spoof the address of the previous web page (the referrer) and trick the server into trusting a malicious payload~\cite{CVE-2011-4909}. Another example~\cite{CVE-2011-4203} is the HTTP response splitting attack which is used to inject malicious payload that could be used to poison the cache of a proxy server that would subsequently serve up the compromised content to unsuspecting clients~\cite{stuttard2008, sullivan2011}. Another form of exploit that involves the HTTP header is called the \emph{MIME sniffing} attack~\cite{CVE-2020-12137}. This is possible because the header contains a content type that specifies the MIME standard (E.g., image/jpg, text/javascript, text/html) of the payload. The browser uses the content type to determine how to treat the message. However, some browsers may determine the content type by examining the payload (sniffing) and rely on this instead of the header. An attacker can leverage this behavior to get the browser to execute malicious code. For instance, suppose the header specifies a content type of image/jpg, but the attacker is able to inject javascript at the top of the image. When the browser sniffs the javascript, it executes code rather than renders an image. This vulnerability can be attributed to not validating MIME types.  We found HTTP header exploits for \emph{Cross-Site Scripting}, \emph{SQL Injection}, \emph{Exposure of Sensitive Information}, and \emph{Code Injection} weaknesses. As Table~\ref{t:MethodVectorMapping} shows, HTTP request is a common attack vector across all types of weaknesses.
    
    \item \textit{STUN Request}: Session Traversal Utilities for Network Address Translation (STUN) is a network protocol for mapping IP addresses between networked resources~\cite{STUN}. Our review showed a vulnerability where the browser disclosed private IP addresses in STUN requests~\cite{CVE-2018-6849}. This type of issue is typically attributed to a lack of encryption.
    
    \item \textit{DNS Request}: The Domain Name System (DNS) provides a mechanism for resolving web addresses (domain names) to IP addresses. The resolutions are accomplished by a client sending queries to a DNS server (nameserver)~\cite{DNS}. The resolved domain name-IP address mapping has a time-to-live (TTL) which upon expiry, a new query is needed to refresh the IP address. The TTL is set by the nameserver. Due to the need for the client to re-resolve domain names (when the TTL expires), an attack-controlled nameserver could swap out IP addresses so that a subsequent query for the same domain resolves to a different IP address. The client application is thus inadvertently used to target a host at the new IP address. This is known as \emph{DNS rebinding} attack, and it allows a compromised page to make cross-domain requests to circumvent the same-origin policy of the HTTP protocol. The same origin policy only permits requests to the same domain (origin) of the page in which the request is included~\cite{stuttard2008, sullivan2011} A DNS rebinding vulnerability was reported and attributed to \emph{Improper Input Validation} weakness~\cite{CVE-2017-2100}. Other similar vulnerabilities involved the client consuming malicious code (in a query response) from a nameserver with a cache that has been poisoned via a DNS cache poisoning attack~\cite{CVE-2014-9509}. Also, we observed a vulnerability where the attacker has utilized DNS requests to bypass two factor authentication~\cite{CVE-2016-10826}.
\end{enumerate}

\begin{enumerate}[label=\roman*., start=5]
    \item \textit{AJAX Request}: Asynchronous JavaScript and XML (AJAX) is a framework that is used to load data into a web page without the need to refresh the entire page. AJAX requests have been used to exploit several weaknesses. CVE-2016-2158~\cite{CVE-2016-2158} is an example that exploits \emph{Exposure of Sensitive Information} weakness. Some AJAX requests utilize \emph{XMLHTTPRequest}. XMLHTTPRequest is form of HTTP request that exchanges XML data between the client and server~\cite{stuttard2008}. As previously discussed, XML data could be vulnerable to \emph{XXE injection}. AJAX at times will use JSON for serialized data transmission, instead of XML. JSON objects (and a variation called JSONP) are processed by the JavaScript interpreter on the client-side. Some vulnerabilities have leveraged this technology to inject malicious code into JSON data to exploit several weaknesses including \emph{SQL Injection} and \emph{XSS}~\cite{CVE-2012-0253}. Our review found vulnerabilities involving AJAX for all types of weakness (Table~\ref{t:MethodVectorMapping}).
    
    \item \textit{RPC Request}: Remote Procedure Call (RPC) is a communication paradigm that allows two independent processes to interact over a network through message-passing~\cite{RPC}. With RPC, a client can remotely invoke methods on a server process (or web service). RPC can be performed over various transport and application layer protocols, including HTTP and SMTP (Simple Mail Transfer Protocol). SOAP (Simple Object Access Protocol) is an XML-based messaging standard for RPC~\cite{bacon2003operating}. Several vulnerabilities involving SOAP requests are reported in the NVD~\cite{CVE-2012-1120}. Some web applications use some form of RPC and at times the interaction includes serialized XML or JSON data. For RPC-based requests, we found it to be associated with all weaknesses except \emph{Path Traversal } (Table~\ref{t:MethodVectorMapping}).
\end{enumerate}

\subsection{Which technologies are susceptible to each weakness?}
Here we identify the specific technologies that were susceptible to the weakness or were utilized in exploiting them. A keyword-based search was utilized, similar to that used in identifying attack vectors. As illustrated in the weakness ontology (Figure~\ref{f:weakness_ontology}), \emph{Technology} is an attribute of each weakness and this attribute has five subtypes. The subtypes are \emph{Language/Framework, Protocol, Data}, \emph{Configuration/Template}, and \emph{Multimedia}. The Data subtype excludes multimedia (i.e., audio, video, and image data). It is important to understand the association between weaknesses and specific technologies as the susceptibility of each technology does vary. Table~\ref{t:MultimediaProtocolMapping}\footnote{See appendix for explanation of acronyms} shows the association between each weakness and their Multimedia attribute, which has 9 subtypes. The subtypes constitute the various multimedia formats that we found associated with each weakness. Our analysis shows that \emph{JP(E)G} (JPG or JPEG image format) played a part in more types of weaknesses than any other multimedia type. CVE-2018-20149~\cite{CVE-2018-20149} is an example of such a vulnerability that relates to XSS weakness.

Understanding weakness as they relate to specific technologies is useful for the requirements engineering of systems~\cite{shostack2014, Goertzel2006}. 
Table~\ref{t:MultimediaProtocolMapping} shows the weaknesses and associated group of attributes that we identified as \emph{Protocol} (with 13 subtypes). It is of no surprise that \emph{HTTP} is common for all weaknesses.
Table~\ref{t:LanguageMapping} shows weaknesses as they relate to programming languages and frameworks. Similar to the Multimedia attribute, this Language/Framework attribute is composed of subtypes (24 subtypes).  Our analysis showed that 5 of these, namely, \emph{AJAX, ASP(X), (P,X)HTML, JavaScript}, and \emph{PHP} can be associated with every type of weakness. Table~\ref{t:DataMapping2} captures the association between weaknesses and the attributes \emph{Data} and \emph{Configuration/Template}. \emph{Data} represents various formats for data encapsulation, of which we found 11 subtypes. \emph{XML, JSON(P)}, and \emph{ZIP} are common data formats for most types of weaknesses.  \emph{Configuration/Template} includes template and configuration files that are used to specify application rules. We identified 7 subtypes for this attribute. \emph{HTACCESS}, the configuration file for Apache Web Server~\cite{ApacheServer}, is the most common of the \emph{Configuration/Template} type.

\begin{table*}[!htb]
\NewDocumentCommand{\rot}{O{45} O{1em} m}{\makebox[#2][l]{\rotatebox{#1}{#3}}}%
  \centering
  \caption{Weaknesses and associated Protocol and Multimedia}
  \label{t:MultimediaProtocolMapping}
  \begin{tabular}{@{}|l|c|c|c|c|c|c|c|c|c|c|c|c|c|c|c|c|c|c|c|c|c|c|@{}}
    \hline
    \textbf{Weakness} & \multicolumn{13}{ c |}{\textbf{Protocol}} & \multicolumn{7}{ c |}{\textbf{Multimedia}} \\ 
\cline{2-21}
    & \rot[90]{\textbf{FTP}}
    & \rot[90]{\textbf{DNS}}
    & \rot[90]{\textbf{HTTP}}
    & \rot[90]{\textbf{LDAP}}
    & \rot[90]{\textbf{MIME}}
    & \rot[90]{\textbf{OAuth}}
    & \rot[90]{\textbf{SMTP}}
    & \rot[90]{\textbf{SOAP}}
    & \rot[90]{\textbf{SSH}}
    & \rot[90]{\textbf{SSL}}
    & \rot[90]{\textbf{STUN}}
    & \rot[90]{\textbf{TLS}}
    & \rot[90]{\textbf{RPC}}
    & \rot[90]{\textbf{ARF/TIFF}}
    & \rot[90]{\textbf{GIF}}
    & \rot[90]{\textbf{JP(E)G}}
    & \rot[90]{\textbf{PNG}}
    & \rot[90]{\textbf{SWF}}
    & \rot[90]{\textbf{SVG}}
    & \rot[90]{\textbf{WRF}}\\
    \hline
    Cross-Site Scripting (XSS) &&& x &  & x &  &  &  &  & x &  &  &x&&& x & x & x &x& \\
    \hline
    SQL Injection & & & x & & &  & & x & x &  &  & &x&&&&&&&\\
    \hline
    Exposure of Sensitive Info. &&& 
    x &  & x &  & x & x &  & x & x & x&x&&&  && x & x &x\\
    \hline
    Cross-Site Request Forgery & && x&  &  &  &  &  &  &  &  &&x&&&&&x&&\\
    \hline
    Improper Access Control & & &x & x&  &  &  &  &  &  & &x&x&&&x&  &  &&\\
    \hline
    Improper Input Validation & x& x&x&  & x &  & x & x & x & x &  &  &x& x &  &x&& x &x&x\\
    \hline
    Path Traversal & & & x& & x &  & & x &  &  &  &&&&&x&x&&&\\
    \hline
    Buffer Overflow& & & x && x &  & x&  &  &  &  &&x&x&&&&&&x\\
    \hline
    Unrestricted Upload of File & x& &x&  & x &  & &  & x &  &  &&x&&x&x&x& x &&\\
    \hline
    Code Injection &x& &x &  & x &  &  & x &  &  &  &&x&&&x&x&  &&\\
    \hline
    Improper Authentication && &x & x &  & x &  &  x&  &  &  &&x&&&&&&&\\
    \hline
  \end{tabular}
\end{table*}

\begin{table*}[!htb]
\NewDocumentCommand{\rot}{O{45} O{1em} m}{\makebox[#2][l]{\rotatebox{#1}{#3}}}%
  \centering
  \caption{Weaknesses and associated Language/Framework}
  \label{t:LanguageMapping}
  \begin{tabular}{@{}|l|c|c|c|c|c|c|c|c|c|c|c|c|c|c|c|c|c|c|c|c|@{}}
    \hline
    \textbf{Weakness} & \multicolumn{20}{ c |}{\textbf{Language/Framework}}  \\ 
\cline{2-21}
    & \rot[90]{\textbf{ActiveX}}
    & \rot[90]{\textbf{AJAX / JavaScript}}
    & \rot[90]{\textbf{AngularJS / VB(Script)}}
    & \rot[90]{\textbf{ASHX}}
    & \rot[90]{\textbf{ASMX}}
    & \rot[90]{\textbf{ASP(X) / PHP}}
    & \rot[90]{\textbf{C\#}}
    & \rot[90]{\textbf{CGI}}
    & \rot[90]{\textbf{CSS}}
    & \rot[90]{\textbf{DLL}}
    & \rot[90]{\textbf{(P,X)HTML}}
    & \rot[90]{\textbf{Java}}
    & \rot[90]{\textbf{jQuery}}
    & \rot[90]{\textbf{JSP(A)}}
    & \rot[90]{\textbf{Perl}}
    & \rot[90]{\textbf{Python}}
    & \rot[90]{\textbf{(re)CAPTCHA}}
    & \rot[90]{\textbf{RDF}}
    & \rot[90]{\textbf{Ruby}}
    & \rot[90]{\textbf{SQL}}\\
    \hline
    Cross-Site Scripting (XSS) &  & x & x & x &  & x & x & x & x & & x & x &x&x&&x&x&&x&x\\
    \hline
    SQL Injection &  & x & & & x & x & x & x &  & & x & x &&x&x&x&&x&x&x\\
    \hline
    Exposure of Sensitive Information &  & x & & &  & x &  & x & x & & x & x &x&x&&x&x&x&x&x\\
    \hline
    Cross-Site Request Forgery &  & x & & x & x & x & x & x &  & & x & x &&x&x&x&&&&\\
    \hline
    Improper Access Control &  & x & & &  & x & & x &  & & x & x &&&&x&x&&&x\\
    \hline
    Improper Input Validation &  & x & & &  & x & x & x & x & x  & x & x&&x&x&x&&&x&x\\
    \hline
    Path Traversal & x & x & & &  & x & x & x &  &  & x & x &&x&x&&x&&&\\
    \hline
    Buffer Overflow & x & x & & &  & x & & &  & x & x & &&&&&&&&\\
    \hline
    Unrestricted Upload of File &  & x & & &  & x & x & x &  &  & x & &&x&&&&&&\\
    \hline
    Code Injection &  & x & & &  & x & & x &  & & x & &&&&x&&&&x\\
    \hline
    Improper Authentication &  & x & & x &  & x & x & x &  &  & x & x &&&&&x&&&\\
    \hline
  \end{tabular}
\end{table*}

\begin{table*}[!htb]
\NewDocumentCommand{\rot}{O{45} O{1em} m}{\makebox[#2][l]{\rotatebox{#1}{#3}}}%
  \centering
  \caption{Weaknesses and associated Data and Configuration/Template}
  \label{t:DataMapping2}
  \begin{tabular}{@{}|l|c|c|c|c|c|c|c|c|c|c|c|c|c|c|c|c|c|c|c|c|@{}}
    \hline
    \textbf{Weakness} & \multicolumn{11}{ c |}{\textbf{Data }} & \multicolumn{7}{ c |}{\textbf{Configuration/Template}} \\ 
\cline{2-19}
    & \rot[90]{\textbf{Atom}}
    & \rot[90]{\textbf{CSV}}
    & \rot[90]{\textbf{DOCX}}
    & \rot[90]{\textbf{JSON(P)}}
    & \rot[90]{\textbf{RSS}}
    & \rot[90]{\textbf{RPT}}
    & \rot[90]{\textbf{TXT}}
    & \rot[90]{\textbf{XLSX}}
    & \rot[90]{\textbf{XML}}
    & \rot[90]{\textbf{YML}}
    & \rot[90]{\textbf{ZIP}}
    & \rot[90]{\textbf{CF}}
    & \rot[90]{\textbf{CG}}
    & \rot[90]{\textbf{CONF}}
    & \rot[90]{\textbf{CTP}}
    & \rot[90]{\textbf{HTACCESS}}
    & \rot[90]{\textbf{TMPL}}
    & \rot[90]{\textbf{TPL}}\\
    \hline
    Cross-Site Scripting (XSS) & x & x & x & x & x &  &  &  & x &  &x&&& x & x & x && x\\
    \hline
    SQL Injection &  &  &  & x & &  &  &  & x & &&&&&&&&\\
    \hline
    Exposure of Sensitive Information & 
    x & x & x & x & x &  & x & x & x & x&x&&& x && x & x &\\
    \hline
    Cross-Site Request Forgery &  & x &  & x & x &  &  &  & x &&x&&&&&&&\\
    \hline
    Improper Access Control &  &  &  & x & x &  &  &  & x &&x&&&&  & x &&\\
    \hline
    Improper Input Validation &  & x &  & x & x &  &  &  & x &  &x& x & x &&& x &&\\
    \hline
    Path Traversal &  &  &  & x & &  &  &  &  &&x&&&&&&&\\
    \hline
    Buffer Overflow&  &  &  &  & & x &  &  & x &&&&&&&&&\\
    \hline
    Unrestricted Upload of File &  & x & x & x & &  & x &  & x &&x&&&&& x &&\\
    \hline
    Code Injection &  & x &  & x & x &  &  &  & x &&x&&&&& x &&\\
    \hline
    Improper Authentication &  &  &  &  & x &  &  &  &  &&x&&&&&&&\\
    \hline
  \end{tabular}
\end{table*}

%% file: s5_RelatedWork.tex
\newpage
\section{Related Work}
\label{s:RelatedWork}
Some works exist that have seek to understand the properties of software vulnerabilities. However, none of those works address the questions posed  by this work with regards to the most common web application weaknesses. Besides, existing works become quickly outdated as the software vulnerability landscape changes. The work presented here involves a review of contemporary vulnerabilities in the NVD. The weakness attributes described are not merely a proposals but based on actual systematic analysis of 10 years (2011 to 2020) of reported vulnerabilities. We are yet to find related work that presents this view.
We organize the related work below into two groups and present a comparison of the most closely related in Section~\ref{s:evaluation}.

\subsection{Web Systems}
\label{s:webVuln}
Sadqi et al.~\cite{Sadqi2020} offers \emph{a comparison} of some existing taxonomies of web application vulnerabilities and threats. It describes the general characteristics of each taxonomy and discusses their advantages and disadvantages. They propose a unified approach that takes advantage of the benefits of the existing taxonomies and classifies attacks by both \emph{client-side} and \emph{server-side}.

Alvarez et al.~\cite{Alvarez2003, Alvarez2003-2} propose a taxonomy of web attacks. Their proposed taxonomy is based on an attack life cycle that includes the following classifiers;(1) Entry point: where the attack gets through; (2) Vulnerability: a weakness in the system that allows the an unauthorized action; (3) Service: the security service threatened by the attack; (4) Action: the attack that realizes the threat (e.g., read, modify, and search); (5) Length: the length of the arguments passed to the HTTP request; (6)  HTTP Verb: Verbs used in the attack (e.g., GET, POST, and HEAD); (7) HTTP Header: the headers used in the attack (e.g., Host, Cookie, and Referer); (8) Target: whether application attacks or platform attacks; (9) Scope: the impact of the attack (local or universal) on the web server users; (9) Privileges: the privileges obtained by the attacker attack (i.e., unprivileged or administrative). This proposed taxonomy is highly generalized and the authors do not offer any basis for deriving their classification criteria.

Al-Kahla et al.~\cite{Wafaa2021} gives \emph{some examples} of web application vulnerabilities which they classify into three type: (1) Vulnerabilities at the input side due to misconfiguration (e.g., broken authentication and session management); (2) Vulnerabilities at the input and output side (e.g., broken access control); (3) vulnerabilities at the output side (e.g., sensitive data exposure). There are also some detection tools for web application vulnerabilities into three types: (1) Static tools for source code review; (2) Dynamic tool detection tools for testing web applications; (3) Hybrid tool that combine both static and dynamic methods.

Lai et al.~\cite{Lai08} propose a taxonomy of Web attacks. The taxonomy is focused on HTTP methods (GET and POST). The given rationale is that attack behavior \emph{only} occurs by transmitting messages between client and server. This assertion is false as some weaknesses such as \emph{Exposure of Sensitive Information} could be present in various system artifacts such as client-side source code. Also, as we show in Table~\ref{t:MultimediaProtocolMapping}, there are other application layer protocols that are susceptible to attack such as FTP, DNS, and LDAP.

Silva et al.~\cite{Silva2016} propose a classification model for vulnerabilities in the web ecosystem. The ecosystem is divided into three domains: (1) Service; (2) Service Consumption; (3) Social Engineering. The classification model is based on 21 attack vectors divided into 8 security threats across the ecosystem domains. Examples of security threats include \emph{Frauds in Service Domain} and \emph{Frauds in Consumer Domain}. Examples of attack vectors include \emph{Missing Cross-Origin Management} and \emph{Service Openness Flaws}. They use an example of the use of their classification by applying it to a list of security risks~\cite{OWASPTopTen} to show their associated attack vectors and threats.

The OWASP (Open Web Application Security Project) Top 10~\cite{OWASPTopTen} is a commonly referenced list of security risks for web applications. There are some similarities between this list and our list of the most common weaknesses in Table~\ref{t:MostWeaknesses}. It is not clear to us how the OWASP Top Ten are determined. The Top 10 is also not strictly based on the CWE~\cite{CWE}, which makes some of its items less concise. For instance \emph{Security Misconfiguration} could be related to several weaknesses including \emph{XSS, CSRF, Improper Access Control, and Exposure of Sensitive Information}. The Top 10 includes \emph{Insufficient Logging \& Monitoring}, which really isn't a software development issue. In addition, it includes \emph{Using Components with Known Vulnerabilities} which is rather broad and not a specific weakness as defined by the CWE. Lastly, the Top 10 does not include \emph{Buffer Overflow} which we did find and is a very common weakness~\cite{Homaei2017, SoftwareType-SEKE-2020, Writing2002, Cert2011}. This work provides a ranking to most common web application weaknesses based solely on the review of 10 years of contemporary vulnerability data. It does not include sundry weaknesses. We view the OWASP Top 10 to be complementary.

Ant\'{o}n et al.~\cite{Anton2004} present a taxonomy of privacy vulnerabilities for Web sites. The classification includes seven kinds of privacy vulnerabilities namely: information monitoring, information aggregation, information storage, information transfer, information collection, information personalization, and contact (of customer). Each vulnerability is associated with privacy goals that can be used to drive privacy policy and reduce privacy vulnerabilities.

Hansman et al.~\cite{Hansman2005} propose a taxonomy of computer network attacks. The taxonomy consists of four dimensions. The first dimension consists of two options: whether the attack uses an attack vector or not. It lists 9 attacks, including Buffer Overflow which it claims does not have an attack vector. The second dimension covers the targets of the attack, which are grouped into hardware and software. The third dimension covers the vulnerabilities and
exploits used in the attack. It claims that entries in this dimension are usually a CVE entry. The fourth dimension relates to attacks having payloads or effects. There are 5 categories to the fourth dimension: (1) First dimension attack payload; (2) Corruption of information; (3) Disclosure of information; (4) Theft of service; (5) Subversion.

Chen et al.~\cite{Chen2018} present a taxonomy of security attacks for IoT. The taxonomy classifies attacks in the four-layer IoT architecture, namely, application layer, middleware layer, network layer, and perception layer. Examples of 23 attacks across all layers are discussed. These attacks include, Code Injections and Buffer overflow in the application layer, SQL Injection and Web browser attack in the middleware layer, DOS attack and Sniffing attack in the network layer, Tag cloning and Eavesdropping in the perception layer. 

Li and Deepa~\cite{Li2014, Deepa2016} put web application vulnerabilities and attacks into 3 groups: (1) Input validation vulnerabilities; (2) Session management vulnerabilities; (3) Application logic vulnerabilities. They identify some examples of vulnerabilities for each group, including, SQL Injection (input validation), Cross-Site Request Forgery (session management), and Redirection Headers (application logic). We believe that this grouping is inaccurate. We found that all common web application weaknesses are susceptible to some form of injection attack, including Cross-Site Request Forgery (Table~\ref{t:MethodVectorMapping}). Although Cross-Site Request Forgery does leverage a user's sessions, some forms of this weakness do exploit the improper validation of input data.

Atashzar et al.~\cite{Atashzar2011} present a review of web application vulnerabilities and hacking tools. The work focuses on the OWASP Top Ten risks 
and its countermeasures released in 2010. The discuss some solutions for the mentioned vulnerabilities. They also present some web application hacking tools and ways to mitigate them.

\subsection{General Software Systems}
Weaver et al.~\cite{Weaver2003} discuss a taxonomy of computer worms. The taxonomy is into 5 categories, namely, target discovery, distribution mechanisms, activation, payloads, and attacker motivation. Target discovery is the mechanism by which a worm discovers its targets. The carrier is the mechanism which a worm uses to transmit to the target. Activation is the mechanism of operation on the target. Payloads are the non-propagating routines a worm utilizes to accomplish its goal. Examples are given in each category. They include human activation (activation), internet DOS (payloads), scanning (target discovery), embedded (carrier), pride and power (motivation).

The MITRE Corporation provides an annual list of the Top 25 most dangerous software weaknesses~\cite{CWETop25-2021}. It determines the \emph{most dangerous} by using a metric that is a function of frequency of occurrence of the weakness and its CVSS score~\cite{CVSS2021}. Although this list leverages NVD data, it is an aggregate of weaknesses as it does not determine weaknesses by software type. It is important to determine common weaknesses by software type (e.g., Operating System and Web Application) as the vulnerability landscape for each type does vary~\cite{SoftwareType-SEKE-2020}. Also, as acknowledged in the limitations of the methodology, the use of this metric could lead to biases in weakness ranking~\cite{CWETop25-2021}.

Tsipenyuk et al.~\cite{Tsipenyuk2005} provide a classification for the common types of coding errors that lead to vulnerabilities. The coding errors are classified into 7 groups: (1) Input Validation and Representation; (2) API Abuse; (3) Security Features; (4) Time and State; (5) Errors; (6) Code Quality; (7) Encapsulation. Some examples of coding errors is given for each group, they include, Log Forging (Input Validation and Representation), Dangerous Function (API Abuse), Insecure Randomness ( Security Features), Deadlock (Time and State), Empty Catch Block (Errors), Double Free (Code Quality), and System Information Leak (Encapsulation).

Li et al.~\cite{Trivedi2017} propose an approach to software vulnerability classification. In their classification, the vulnerabilities are divided into three categories: Aging related vulnerability, Non aging Mandel Vulnerability, and Bohr Vulnerability. They identify 7 attack patterns across all three categories. The attack patterns are, race condition attack, TOCTOU attack, use after free attack, integer overflow attack, type confusion attack, uninitialization attack, and buffer overflow attack.

Almutairy et al.~\cite{Almutairy2019} present a taxonomy of virtualization security challenges. The challenges are grouped into 6 categories. The categories are characteristics-related issues, infrastructure issues, access issues, data security issues, control and monitoring issues, and security policy and rules. A total of 30 vulnerabilities are listed across all 6 categories. Examples of the vulnerabilities in each category include, Incorrect Isolation, Insecure Hypervisor, Hidden Identity, Improper Data Sanitization, Lack of Visibility Control, and Lack of Security Policy.

Piessens~\cite{Piessens2002} \emph{proposes} a taxonomy of the most frequently occurring causes of vulnerability. The taxonomy is organized into five categories. The categories are analysis phase, design phase, implementation phase, deployment phase, and maintenance phase. It identifies 19 causes of vulnerabilities across all the phases. Examples of the causes include, no risk analysis in analysis phase, no logging in design phase, insufficiently defensive input checking in implementation phase, complex or unnecessary configuration in deployment phase, and insecure fallback in the maintenance phase.

Berghe et al.~\cite{Berghe05} propose a methodology for predicting vulnerabilities in systems. The methodology which assesses the likelihood that historic vulnerabilities will appear in a system, is based on determining the properties of the system (e.g.,use of encryption) and its architecture. They apply their methodology by utilizing an example of a web service. In the example, a matrix (weighted) is used to predict the likelihood of a vulnerability by mapping some vulnerabilities (e.g., configuration error, design error) against some architectural properties (e.g., client, directory, and backend). The system properties, vulnerabilities, and their weighting as used in this methodology, seem arbitrary.

Igure et al.~\cite{Igure2008} present an overview of taxonomies of attacks and vulnerabilities from 1974 until 2006 is presented. Their work is similar to that of Sadqi et al.~\cite{Sadqi2020} above. It presents a survey of taxonomies that are related to computer and network security. They claim that their comparison of taxonomies helps identify specific properties of that aid security assessment.

\subsection{Evaluation against related works}
\label{s:evaluation}
In this section, we offer a comparison of this work against related work. We focus the comparison on the most related, which are the web systems vulnerabilities (Section~\ref{s:webVuln}). We use this comparison to demonstrate the contribution of this work. We evaluate these works against the criteria listed below. These criteria are aligned with the research questions (see Section~\ref{s:Introduction}) addressed by this work.

\begin{enumerate}[label=(\Alph*)]
    \item The work quantifies the most reported web application vulnerabilities?
    \item The work enumerates the most common weaknesses for web applications?
    \item The work enumerates the attack methods that are utilized to exploit each weakness?
    \item The work enumerates the attack vectors that were leveraged to exploiting each weakness?
    \item The work enumerates the technologies that are susceptible to each weakness?
\end{enumerate}

Table~\ref{t:TaxonomyAnalysis} summarizes the comparison of closely related work. As can be seen, no other work meets all the 5 criteria listed above, as indicated by the check marks. This work is the only one that performs a comprehensive review of vulnerability data to determine the most common web applications weaknesses. And based on that, provides a weakness-specific classification that maps each weakness to their attack method, attack vectors, and impacted technologies, as determined by vulnerability data. Similar works are not comprehensive and merely use examples vulnerabilities to derive their classification.

\begin{table*}[!htb]
\NewDocumentCommand{\rot}{O{45} O{1em} m}{\makebox[#2][l]{\rotatebox{#1}{#3}}}%
  \centering
  \caption{Comparison of related works.}
  \label{t:TaxonomyAnalysis}
  \begin{tabular}{@{}|l|l|c|c|c|c|c|l|@{}}
    \hline
    \textbf{Work} & \textbf{Goal} & \multicolumn{5}{ c |}{\textbf{Criteria}} & \multicolumn{1}{ c |}{\textbf{Comment}}\\ 
\cline{3-8}
    &
    & \rot[0]{\textbf{A}}
    & \rot[0]{\textbf{B}}
    & \rot[0]{\textbf{C}}
    & \rot[0]{\textbf{D}}
    & \rot[0]{\textbf{E}}
    & \\
    \hline
    Sadqi et al.~\cite{Sadqi2020} &\Centerstack[l]{Compare some existing vulnerability \\classifications.}&\xmark& \xmark & \checkmark & \checkmark & \checkmark & \Centerstack[l]{Offers summaries of different taxonomies.\\Some of which are presented here.}\\
    \hline
    Alvarez et al.~\cite{Alvarez2003, Alvarez2003-2} &Classification of web attacks.&\xmark&\xmark& \checkmark & \checkmark & \checkmark &  \Centerstack[l]{Focuses only on "HTTP attacks".\\
    Identifies 5 vulnerabilities (Code injection,\\ Canonicalization, HTMLmanipulation, \\ Overflow, and Misconfiguration)}\\
    \hline
    Al-Kahla et al.~\cite{Wafaa2021} &Classify web application vulnerabilities.&\xmark&\xmark& \checkmark & \checkmark & \checkmark & \Centerstack[l]{Groups vulnerabilities into 3 broad types (Input \\side, Input \& Output side, and Output side)\\Gives some examples of each type but only \\
   SQL injection is listed as "injection flaws".}\\
    \hline
    Lai et al.~\cite{Lai08}&Classification of web attacks.&\xmark&\xmark&\checkmark&\checkmark&\checkmark&\Centerstack[l]{Focuses on attacks on "HTTP methods".\\
    Identifies 5 types of attacks \\(SQL injection, Buffer overflow, Access \\control, Authorization, and Authentication)}\\
    \hline
    Silva et al.~\cite{Silva2016} &\Centerstack[l]{A vulnerability classification model for \\web ecosystem.}&\xmark& \xmark & \checkmark & \checkmark & \xmark & \Centerstack[l]{Proposed taxonomy is highly generalized and \\not vulnerability-centric.}\\
    \hline
    OWASP~\cite{OWASPTopTen} &Identify top 10 web application risks.&\xmark& \checkmark & \checkmark & \checkmark & \checkmark & \Centerstack[l]{Does not offer classification for vulnerabilities.\\Ranking not based solely on vulnerability data.}\\
    \hline
    Ant\'{o}n et al.~\cite{Anton2004} &\Centerstack[l]{Classify web site privacy vulnerabilities.\\}&\xmark& \xmark & \xmark & \xmark & \xmark & Focuses on privacy requirements.\\
    \hline
    Hansman et al.~\cite{Hansman2005} &\Centerstack[l]{Classify computer network attacks.\\}&\xmark&\xmark & \xmark & \checkmark & \checkmark & Focuses on viruses and worms.\\
    \hline
    Chen et al.~\cite{Chen2018} &Classify vulnerabilities in the IoT.&\xmark& \xmark & \checkmark& \checkmark & \checkmark &\Centerstack[l]{Gives some examples of vulnerabilities in 4\\architecural layers (application, middleware, \\network, and perception).}\\
    \hline
    Li et al.~\cite{Li2014} &\Centerstack[l]{Discuss server-side vulnerability \\mitigation techniques.}&\xmark& \xmark & \checkmark & \checkmark & \checkmark &\Centerstack[l]{Groups vulnerabilities into 3 types (input \\validation, session management and
application \\logic).\\Leverages OWASP Top 10 risks.}\\
    \hline
    Deepa et al.~\cite{Deepa2016} &\Centerstack[l]{Discuss web application vulnerability \\mitigation techniques.}&\xmark& \xmark & \checkmark & \checkmark & \checkmark &\Centerstack[l]{Groups vulnerabilities into 3 types (improper input \\validation, improper authentication and \\authorization, and
Improper session management).\\Gives some examples of each type.}\\
\hline
    Atashzar et al.~\cite{Atashzar2011}&\Centerstack[l]{Discuss web application vulnerabilities. \\}&\xmark& \checkmark & \checkmark & \checkmark & \checkmark &\Centerstack[l]{Leverages OWASP Top 10 risks.\\}\\
    \hline
  \end{tabular}
\end{table*}

%% file: s6_Conclusion.tex
\section{Discussion and Conclusion}
\label{s:Conclusion}
\balance 

To help address the proliferation of web application vulnerabilities, this work presents a review of contemporary vulnerability data. The data from the NVD covers the period of 2011 to 2020. With the review, we address 6 research questions to understand the characteristics of web application weakness. The 11 most common web application weaknesses are identified and their properties are captured as a set of weakness attributes. These attributes are modeled as a weakness ontology. The ontology describes the relationship between weakness and attack methods, attack vectors, impact, and technologies that are common with each weakness. This ontology will help inform efforts related to threat modeling and security requirements engineering, where more automated tools are needed. This work shows that the most prevalent attack method is \emph{injection}. Instances of this method of attack were found for all types of weaknesses. Similarly, \emph{form data}, \emph{file data}, and \emph{HTTP request} were the most common attack vectors across all weaknesses. We also show that \emph{security} is the most impacted software quality attribute. A wide range of compromised technologies is also identified. These technologies span \emph{languages, frameworks, protocols, multimedia}, and \emph{data} formats.

We note some limitations of NVD data. Both the data and its schema are subject to change. The data used in this review was retrieved in January of 2021. At the time, the schema was \emph{NVD JSON 1.0}. As of this writing, the current version is \emph{NVD JSON 1.1}~\cite{NVDFeed}. With the change in the schema, certain information have changed. For instance, software vendor is no longer captured as a separate element. Also, the vulnerability data does get updated, this is especially true for more recent vulnerabilities. Whereas, vulnerabilities from previous years are more stable. The one-to-one mapping of vulnerabilities to weaknesses that was used in the version of the reviewed data may lead to incorrect count of weaknesses. This is because certain weaknesses are related. For instance, \emph{Improper Access Control} and \emph{Improper Authentication} may be related for a specific vulnerability. However, only one weakness is typically identified for the vulnerability. This one-to-one mapping also leads to missed causal relationships between weaknesses. Such as that which exists between \emph{SQL Injection} and \emph{Improper Input Validation}.

The vulnerability descriptions in each CVE entry do vary in their level of detail. They often do contain a wealth of useful information but this is not always the case. For instance, for some vulnerabilities, the weakness is not identified but rather is noted as \emph{NVD-CWE-noinfo}. A smart software tool to automate the analysis of NVD data in a scalable manner is needed. This tool could utilize machine learning to address some of the limitations in NVD such as identify missing information such as software types, and weaknesses. This tool could also identify causal relationships between weaknesses. The wealth of information available in the NVD presents an opportunity to better understand the software vulnerability landscape, and improve threat modeling and requirements engineering for software.

%% file: s7_Appendix.tex
{\appendix[]
\label{s:Appendix}
ActiveX - Microsoft Software Framework

AngularJS - JavaScript Framework

AJAX - Asynchronous JavaScript and XML

ARF - Advanced Recording Format

ASHX - Web application (.NET Framework)

ASMX - Web Service (.NET Framework)

ASP(X) - (ASP or ASPX)

ASP - Active Server Pages

ASPX - Active Server Pages (.Net Framework)

Atom - A syndication data feed format

C\# - C-sharp Programming Language (.Net Framework)

CF -  Configuration file for Sendmail application

CG - Security policy file for cPanel application

CGI - Common Gateway Interface

CONF - A commonly used configuration file extension

CSS - Cascading Style Sheets

CSV - Comma Separated Value

CTP - Layout template file for MISP application


DLL - Dynamic Link Library

DNS - Domain Name System

DOCX - Microsoft Word Format


FTP - File Transfer Protocol

GIF - Graphics Interchange Format

HTACCESS - Configuration file for Apache Web Server

HTML - Hypertext Markup Language

HTTP - Hypertext Transfer Protocol

jQuery - JavaScript Library

JP(E)G - (JPEG or JPG) image file format

JSON(P) - (JSON or JSONP) JavaScript Object Notation

JSP(A) - (JSP or JSPA) Java Server Pages

LDAP - Lightweight Directory Access Protocol

MIME - Multipurpose Internet Mail Extensions

OAuth - Open Authorization protocol

PHP - Hypertext Preprocessor

PHTML - HTML with PHP

PNG - Portable Network Graphics

(P,X)HTML - PHTML, XHTML, or HTML

(RE)CAPTCHA - Test for determining human users 

RDF - Resource Description Framework

RPC - Remote Procedure Call

RPT - Crystal Reports application file

RSS - Really Simple Syndication

SMTP - Simple Mail Transfer Protocol

SOAP - Simple Object Access Protocol

SQL - Structured Query Language

SSH - Secure Shell Protocol

SSL - Secure Sockets Layer

STUN - Session Traversal Utilities for NAT

SVG - Scalable Vector Graphics

SWF - Adobe Flash file format

TIFF - Tag Image File Format

TMPL - Template file for Bugzilla application

TLS - Transport Layer Security

TPL - A commonly used template file extension

TXT - Text file format

URL - Uniform Resource Locator

VB(Script) - (VB or VBScript) Visual Basic

WRF -  Webex Recording Format

XHTML - Extensible Hypertext Markup Language

XLSX - Microsoft Excel Format

XML - Extensible Markup Language

YAML - Yet Another Markup Language

ZIP - Document archive
}



%% file: bare_jrnl.bbl
\begin{thebibliography}{10}
\providecommand{\url}[1]{#1}
\csname url@samestyle\endcsname
\providecommand{\newblock}{\relax}
\providecommand{\bibinfo}[2]{#2}
\providecommand{\BIBentrySTDinterwordspacing}{\spaceskip=0pt\relax}
\providecommand{\BIBentryALTinterwordstretchfactor}{4}
\providecommand{\BIBentryALTinterwordspacing}{\spaceskip=\fontdimen2\font plus
\BIBentryALTinterwordstretchfactor\fontdimen3\font minus
  \fontdimen4\font\relax}
\providecommand{\BIBforeignlanguage}[2]{{%
\expandafter\ifx\csname l@#1\endcsname\relax
\typeout{** WARNING: IEEEtran.bst: No hyphenation pattern has been}%
\typeout{** loaded for the language `#1'. Using the pattern for}%
\typeout{** the default language instead.}%
\else
\language=\csname l@#1\endcsname
\fi
#2}}
\providecommand{\BIBdecl}{\relax}
\BIBdecl

\bibitem{NVD2021}
``National {Vulnerability Database},'' \url{https://nvd.nist.gov/}, Retrieved
  March, 2021.

\bibitem{Goertzel2006}
K.~Goertzel, T.~Winograd, H.~McKinley, and P.~Holley, ``Security in the
  software lifecycle: Making software development processes and the software
  produced by them more secure,'' \emph{{U.S. Department of Homeland
  Security}}, August 2006.

\bibitem{Tuma2018}
K.~Tuma, G.~Calikli, and R.~Scandariato, ``\BIBforeignlanguage{English}{Threat
  analysis of software systems: A systematic literature review},''
  \emph{\BIBforeignlanguage{English}{Journal of Systems and Software}}, vol.
  144, Oct. 2018.

\bibitem{Writing2002}
M.~Howard and D.~LeBlanc, \emph{Writing Secure Code}.\hskip 1em plus 0.5em
  minus 0.4em\relax Microsoft Press, 2002.

\bibitem{OOSE2009}
B.~Bruegge and A.~H. Dutoit, \emph{Object-Oriented Software Engineering Using
  {UML}, Patterns, and Java}, 3rd~ed.\hskip 1em plus 0.5em minus 0.4em\relax
  Prentice Hall Press, 2009.

\bibitem{Cox2019}
R.~Cox, ``Surviving software dependencies,'' \emph{Communications of the
  {ACM}}, vol.~62, no.~9, August 2019.

\bibitem{Ivaki2015}
N.~Ivaki, N.~Laranjeiro, and F.~Araujo, ``A taxonomy of reliable
  request-response protocols,'' in \emph{Proceedings of the 30th Annual ACM
  Symposium on Applied Computing}.\hskip 1em plus 0.5em minus 0.4em\relax New
  York, NY, USA: Association for Computing Machinery, 2015.

\bibitem{guinard2011}
D.~Guinard, V.~Trifa, F.~Mattern, and E.~Wilde, ``From the internet of things
  to the web of things: Resource-oriented architecture and best practices,'' in
  \emph{Architecting the Internet of things}.\hskip 1em plus 0.5em minus
  0.4em\relax Springer, 2011.

\bibitem{hadley2006}
M.~J. Hadley, ``Web application description language {WADL},'' 2006.

\bibitem{guinard2011web}
D.~Guinard, ``A web of things application architecture: Integrating the
  real-world into the web,'' Ph.D. dissertation, ETH Zurich, 2011.

\bibitem{SoftwareType-SEKE-2020}
O.~Ezenwoye, Y.~Liu, and W.~Patten, ``Classifying common security
  vulnerabilities by software type,'' in \emph{International Conference on
  Software Engineering and Knowledge Engineering}, 2020.

\bibitem{Ezenwoye2022}
O.~Ezenwoye and Y.~Liu, ``Integrating vulnerability risk into the software
  process,'' in \emph{Proceedings of the 2022 ACM Southeast Conference}.\hskip
  1em plus 0.5em minus 0.4em\relax Association for Computing Machinery, 2022.

\bibitem{Lee21}
D.~Lee, B.~Steed, Y.~Liu, and O.~Ezenwoye, ``Tutorial: A lightweight web
  application for software vulnerability demonstration,'' in \emph{2021 {IEEE}
  Secure Development Conference}.\hskip 1em plus 0.5em minus 0.4em\relax {IEEE}
  Computer Society, October 2021.

\bibitem{CVEProgram}
``{CVE} program,'' \url{https://www.cve.org/About/Overview}, Retrieved
  {December}, 2021.

\bibitem{CWE}
``{Common Weakness Enumeration},'' \url{https://nvd.nist.gov/vuln/categories},
  Retrieved {March}, 2021.

\bibitem{Exploiting2004}
G.~Hoglund and G.~McGraw, \emph{Exploiting Software: How to Break Code}.\hskip
  1em plus 0.5em minus 0.4em\relax Pearson Higher Education, 2004.

\bibitem{Howto2006}
M.~Andrews and J.~A. Whittaker, \emph{How to Break Web Software: Functional and
  Security Testing of Web Applications and Web Services}.\hskip 1em plus 0.5em
  minus 0.4em\relax Addison-Wesley Professional, 2006.

\bibitem{stuttard2008}
D.~Stuttard and M.~Pinto, \emph{The Web Application Hacker's Handbook:
  Discovering and Exploiting Security Flaws}.\hskip 1em plus 0.5em minus
  0.4em\relax Wiley, 2008.

\bibitem{sullivan2011}
B.~Sullivan and V.~Liu, \emph{Web Application Security, A Beginner's
  Guide}.\hskip 1em plus 0.5em minus 0.4em\relax McGraw-Hill Education, 2011.

\bibitem{TheArt2006}
M.~Dowd, J.~McDonald, and J.~Schuh, \emph{The Art of Software Security
  Assessment: Identifying and Preventing Software Vulnerabilities}.\hskip 1em
  plus 0.5em minus 0.4em\relax Addison-Wesley Professional, 2006.

\bibitem{Secure2007}
B.~Chess and J.~West, \emph{Secure Programming with Static Analysis},
  1st~ed.\hskip 1em plus 0.5em minus 0.4em\relax Addison-Wesley Professional,
  2007.

\bibitem{Enterprise2014}
K.~R. van Wyk, M.~G. Graff, D.~S. Peters, and D.~L. Burley, \emph{Enterprise
  Software Security: A Confluence of Disciplines}, 1st~ed.\hskip 1em plus 0.5em
  minus 0.4em\relax Addison-Wesley Professional, 2014.

\bibitem{CWE-287}
``Improper authentication,''
  \url{https://cwe.mitre.org/data/definitions/287.html}, Retrieved {August},
  2021.

\bibitem{VulnViz2021}
{NIST}, ``Vulnerability visualizations,''
  \url{https://nvd.nist.gov/General/Visualizations/Vulnerability-Visualizations},
  Retrieved {June}, 2021.

\bibitem{CVEDetails2021}
S.~Özkan, ``{CVE} details,'' \url{https://www.cvedetails.com/}, Retrieved
  {June}, 2021.

\bibitem{Homaei2017}
H.~Homaei and H.~R. Shahriari, ``Seven years of software vulnerabilities: The
  ebb and flow,'' \emph{IEEE Security Privacy}, vol.~15, no.~1, 2017.

\bibitem{Forward2008}
A.~Forward and T.~Lethbridge, ``A taxonomy of software types to facilitate
  search and evidence-based software engineering,'' in \emph{Proceedings of the
  2008 Conference of the Center for Advanced Studies}, 2008.

\bibitem{NVDData2021}
NIST, ``{NVD} data feeds,'' \url{https://nvd.nist.gov/vuln/data-feeds},
  Retrieved {November}, 2021.

\bibitem{dierenfeld2012}
H.~Dierenfeld and A.~Merceron, ``Learning analytics with excel pivot tables,''
  \emph{Moodle Research Conference}, 2012.

\bibitem{michahelles2015}
F.~Michahelles and S.~Mayer, ``Toward a web of systems,'' \emph{XRDS:
  Crossroads, The ACM Magazine for Students}, vol.~22, no.~2, 2015.

\bibitem{shostack2014}
A.~Shostack, \emph{Threat Modeling: Designing for Security}.\hskip 1em plus
  0.5em minus 0.4em\relax Wiley, 2014.

\bibitem{MELLADO2010}
D.~Mellado, C.~Blanco, L.~E. Sánchez, and E.~Fernández-Medina, ``A systematic
  review of security requirements engineering,'' \emph{Computer Standards \&
  Interfaces}, vol.~32, no.~4, 2010.

\bibitem{Souag2016}
A.~Souag, R.~Mazo, C.~Salinesi, and I.~Comyn-Wattiau, ``Reusable knowledge in
  security requirements engineering: a systematic mapping study,''
  \emph{Requir. Eng.}, vol.~21, no.~2, jun 2016.

\bibitem{CVE-2011-0774}
``Cve-2011-0774,'' \url{https://nvd.nist.gov/vuln/detail/CVE-2011-0774},
  Retrieved {November}, 2021.

\bibitem{CVE-2012-0792}
``Cve-2012-0792,'' \url{https://nvd.nist.gov/vuln/detail/CVE-2012-0792},
  Retrieved {November}, 2021.

\bibitem{CVE-2015-0943}
``Cve-2015-0943,'' \url{https://nvd.nist.gov/vuln/detail/CVE-2015-0943},
  Retrieved {December}, 2021.

\bibitem{Simmons14}
C.~Simmons, S.~Shiva, H.~Bedi, and D.~Dasgupta, ``{AVOIDIT}: A cyber attack
  taxonomy,'' in \emph{9th Annual Symposium on Information Assurance}, June
  2014.

\bibitem{CVE-2020-12432}
``Cve-2020-12432,'' \url{https://nvd.nist.gov/vuln/detail/CVE-2020-12432},
  Retrieved {May}, 2022.

\bibitem{CVE-2019-9714}
``Cve-2019-9714,'' \url{https://nvd.nist.gov/vuln/detail/CVE-2019-9714},
  Retrieved {December}, 2021.

\bibitem{CVE-2020-13241}
``Cve-2020-13241,'' \url{https://nvd.nist.gov/vuln/detail/CVE-2020-13241},
  Retrieved {November}, 2021.

\bibitem{CVE-2017-12216}
``Cve-2017-12216,'' \url{https://nvd.nist.gov/vuln/detail/CVE-2017-12216},
  Retrieved {November}, 2021.

\bibitem{CVE-2017-18123}
``Cve-2017-18123,'' \url{https://nvd.nist.gov/vuln/detail/CVE-2017-18123},
  Retrieved {November}, 2021.

\bibitem{CVE-2017-1000496}
``Cve-2017-1000496,'' \url{https://nvd.nist.gov/vuln/detail/CVE-2017-1000496},
  Retrieved {November}, 2021.

\bibitem{CVE-2020-11807}
``Cve-2020-11807,'' \url{https://nvd.nist.gov/vuln/detail/CVE-2020-11807},
  Retrieved {November}, 2021.

\bibitem{CVE-2017-12979}
``Cve-2017-12979,'' \url{https://nvd.nist.gov/vuln/detail/CVE-2017-12979},
  Retrieved {November}, 2021.

\bibitem{CVE-2017-17727}
``Cve-2017-17727,'' \url{https://nvd.nist.gov/vuln/detail/CVE-2017-17727},
  Retrieved {November}, 2021.

\bibitem{CVE-2016-3002}
``Cve-2016-3002,'' \url{https://nvd.nist.gov/vuln/detail/CVE-2016-3002},
  Retrieved {November}, 2021.

\bibitem{CVE-2015-1951}
``Cve-2015-1951,'' \url{https://nvd.nist.gov/vuln/detail/CVE-2015-1951},
  Retrieved {November}, 2021.

\bibitem{CVE-2015-5687}
``Cve-2015-5687,'' \url{https://nvd.nist.gov/vuln/detail/CVE-2015-5687},
  Retrieved {December}, 2021.

\bibitem{CVE-2014-1695}
``Cve-2014-1695,'' \url{https://nvd.nist.gov/vuln/detail/CVE-2014-1695},
  Retrieved {November}, 2021.

\bibitem{CVE-2017-8385}
``Cve-2017-8385,'' \url{https://nvd.nist.gov/vuln/detail/CVE-2017-8385},
  Retrieved {November}, 2021.

\bibitem{CVE-2018-15408}
``Cve-2018-15408,'' \url{hhttps://nvd.nist.gov/vuln/detail/CVE-2018-15408},
  Retrieved {November}, 2021.

\bibitem{HTTP}
``Http,'' \url{https://datatracker.ietf.org/doc/html/rfc2616}, Retrieved
  {December}, 2021.

\bibitem{CVE-2019-11489}
``Cve-2019-11489,'' \url{https://nvd.nist.gov/vuln/detail/CVE-2019-11489},
  Retrieved {December}, 2021.

\bibitem{CVE-2013-2713}
``Cve-2013-2713,'' \url{https://nvd.nist.gov/vuln/detail/CVE-2013-2713},
  Retrieved {December}, 2021.

\bibitem{CVE-2011-4737}
``Cve-2011-4737,'' \url{https://nvd.nist.gov/vuln/detail/CVE-2011-4737},
  Retrieved {December}, 2021.

\bibitem{CVE-2011-4203}
``Cve-2011-4203,'' \url{https://nvd.nist.gov/vuln/detail/CVE-2011-4203},
  Retrieved {November}, 2021.

\bibitem{CVE-2011-4909}
``Cve-2011-4909,'' \url{https://nvd.nist.gov/vuln/detail/CVE-2011-4909},
  Retrieved {December}, 2021.

\bibitem{CVE-2020-12137}
``Cve-2020-12137,'' \url{https://nvd.nist.gov/vuln/detail/CVE-2020-12137},
  Retrieved {December}, 2021.

\bibitem{STUN}
``{Session Traversal Utilities for NAT},''
  \url{https://datatracker.ietf.org/doc/html/rfc5389}, Retrieved {August},
  2021.

\bibitem{CVE-2018-6849}
``Cve-2018-6849,'' \url{https://nvd.nist.gov/vuln/detail/CVE-2018-6849},
  Retrieved {December}, 2021.

\bibitem{DNS}
``{Domain Name System},'' \url{https://datatracker.ietf.org/doc/html/rfc1034},
  Retrieved {August}, 2021.

\bibitem{CVE-2017-2100}
``Cve-2017-2100,'' \url{https://nvd.nist.gov/vuln/detail/CVE-2017-2100},
  Retrieved {December}, 2021.

\bibitem{CVE-2014-9509}
``Cve-2014-9509,'' \url{https://nvd.nist.gov/vuln/detail/CVE-2014-9509},
  Retrieved {December}, 2021.

\bibitem{CVE-2016-10826}
``Cve-2016-10826,'' \url{https://nvd.nist.gov/vuln/detail/CVE-2016-10826},
  Retrieved {December}, 2021.

\bibitem{CVE-2016-2158}
``Cve-2016-2158,'' \url{https://nvd.nist.gov/vuln/detail/CVE-2016-2158},
  Retrieved {December}, 2021.

\bibitem{CVE-2012-0253}
``Cve-2012-0253,'' \url{https://nvd.nist.gov/vuln/detail/CVE-2012-0253},
  Retrieved {December}, 2021.

\bibitem{RPC}
``Remote procedure call protocol,''
  \url{https://datatracker.ietf.org/doc/html/rfc1831.html}, Retrieved
  {December}, 2021.

\bibitem{bacon2003operating}
J.~Bacon and T.~Harris, \emph{Operating Systems: Concurrent and Distributed
  Software Design}.\hskip 1em plus 0.5em minus 0.4em\relax Addison-Wesley,
  2003.

\bibitem{CVE-2012-1120}
``Cve-2012-1120,'' \url{https://nvd.nist.gov/vuln/detail/CVE-2012-1120},
  Retrieved {December}, 2021.

\bibitem{CVE-2018-20149}
``Cve-2018-20149,'' \url{https://nvd.nist.gov/vuln/detail/CVE-2018-20149},
  Retrieved {December}, 2021.

\bibitem{ApacheServer}
``Apache {HTTP} server project,'' \url{https://httpd.apache.org/}, Retrieved
  {December}, 2021.

\bibitem{Sadqi2020}
Y.~Sadqi and M.~Yassine, ``A systematic review and taxonomy of web applications
  threats,'' \emph{Information Security Journal A Global Perspective}, 12 2020.

\bibitem{Alvarez2003}
G.~{\'A}lvarez and S.~Petrovi{\'{c}}, ``A taxonomy of web attacks,'' in
  \emph{Web Engineering}.\hskip 1em plus 0.5em minus 0.4em\relax Berlin,
  Heidelberg: Springer Berlin Heidelberg, 2003.

\bibitem{Alvarez2003-2}
G.~{\'A}lvarez and S.~Petrovi{\'c}, ``A new taxonomy of web attacks suitable
  for efficient encoding,'' \emph{Computers \& Security}, vol.~22, 07 2003.

\bibitem{Wafaa2021}
W.~Al-Kahla, A.~S. Shatnawi, and E.~Taqieddin, ``A taxonomy of web security
  vulnerabilities,'' in \emph{2021 12th International Conference on Information
  and Communication Systems}, 2021.

\bibitem{Lai08}
J.-Y. Lai, J.-S. Wu, S.-J. Chen, C.-H. Wu, and C.-H. Yang, ``Designing a
  taxonomy of web attacks,'' in \emph{2008 International Conference on
  Convergence and Hybrid Information Technology}, 2008.

\bibitem{Silva2016}
C.~Silva, R.~Batista, R.~Queiroz, V.~Garcia, J.~Silva, D.~Gatti, R.~Assad,
  L.~Nascimento, K.~Brito, and P.~Miranda, ``Towards a taxonomy for security
  threats on the web ecosystem,'' in \emph{2016 IEEE/IFIP Network Operations
  and Management Symposium}, 2016.

\bibitem{OWASPTopTen}
``{OWASP} top 10 web application security risks,''
  \url{https://owasp.org/www-project-top-ten/}, Retrieved April, 2021.

\bibitem{Cert2011}
F.~Long, D.~Mohindra, R.~C. Seacord, D.~F. Sutherland, and D.~Svoboda,
  \emph{The {CERT} Oracle Secure Coding Standard for Java}, 1st~ed.\hskip 1em
  plus 0.5em minus 0.4em\relax Addison-Wesley Professional, 2011.

\bibitem{Anton2004}
I.~Ant\'{o}n and B.~Earp, ``A requirements taxonomy for reducing web site
  privacy vulnerabilities,'' \emph{Requirements Engineering}, vol.~9, no.~3,
  August 2004.

\bibitem{Hansman2005}
S.~Hansman and R.~Hunt, ``A taxonomy of network and computer attacks,''
  \emph{Computers \& Security}, vol.~24, 02 2005.

\bibitem{Chen2018}
K.~Chen, S.~Zhang, Z.~Li, Y.~Zhang, Q.~Deng, S.~Ray, and Y.~Jin,
  ``Internet-of-things security and vulnerabilities: Taxonomy, challenges, and
  practice,'' \emph{Journal of Hardware and Systems Security}, vol.~2, no.~2,
  06 2018.

\bibitem{Li2014}
X.~Li and Y.~Xue, ``A survey on server-side approaches to securing web
  applications,'' \emph{ACM Comput. Surv.}, vol.~46, no.~4, Mar. 2014.

\bibitem{Deepa2016}
G.~Deepa and P.~S. Thilagam, ``Securing web applications from injection and
  logic vulnerabilities,'' \emph{Inf. Softw. Technol.}, vol.~74, no.~C, Jun.
  2016.

\bibitem{Atashzar2011}
H.~Atashzar, A.~Torkaman, M.~Bahrololum, and M.~H. Tadayon, ``A survey on web
  application vulnerabilities and countermeasures,'' in \emph{2011 6th
  International Conference on Computer Sciences and Convergence Information
  Technology}, 2011.

\bibitem{Weaver2003}
N.~Weaver, V.~Paxson, S.~Staniford, and R.~Cunningham, ``A taxonomy of computer
  worms,'' in \emph{Proceedings of the 2003 ACM Workshop on Rapid Malcode},
  2003.

\bibitem{CWETop25-2021}
``2021 {CWE} top 25 most dangerous software weaknesses,''
  \url{https://cwe.mitre.org/top25/archive/2021/2021_cwe_top25.html}, Retrieved
  {November}, 2021.

\bibitem{CVSS2021}
``The common vulnerability scoring system,''
  \url{https://nvd.nist.gov/vuln-metrics/cvss}, Retrieved November, 2021.

\bibitem{Tsipenyuk2005}
K.~Tsipenyuk, B.~Chess, and G.~McGraw, ``Seven pernicious kingdoms: A taxonomy
  of software security errors,'' \emph{IEEE Security and Privacy}, vol.~3,
  no.~6, Nov. 2005.

\bibitem{Trivedi2017}
X.~{Li}, X.~{Chang}, J.~A. {Board}, and K.~S. {Trivedi}, ``A novel approach for
  software vulnerability classification,'' in \emph{2017 Annual Reliability and
  Maintainability Symposium}, 2017.

\bibitem{Almutairy2019}
N.~Almutairy, K.~Al-Shqeerat, and H.~AlHamad, ``A taxonomy of virtualization
  security issues in cloud computing environments,'' \emph{Indian Journal of
  Science and Technology}, vol.~12, p.~19, 01 2019.

\bibitem{Piessens2002}
F.~Piessens, ``A taxonomy of causes of software vulnerabilities in internet
  software,'' in \emph{The 13th International Symposium on Software Reliability
  Engineering}, 2002.

\bibitem{Berghe05}
C.~V. Berghe, C.~V, E.~Berghe, J.~Riordan, and F.~Piessens, ``A vulnerability
  taxonomy methodology applied to web services,'' in \emph{The 10th Nordic
  Workshop on Secure IT Systems}, 2005.

\bibitem{Igure2008}
V.~M. {Igure} and R.~D. {Williams}, ``Taxonomies of attacks and vulnerabilities
  in computer systems,'' \emph{IEEE Communications Surveys Tutorials}, vol.~10,
  no.~1, pp. 6--19, 2008.

\bibitem{NVDFeed}
``{NVD} data feeds,'' \url{https://nvd.nist.gov/vuln/data-feeds}, Retrieved
  {December}, 2021.

\end{thebibliography}
